\documentclass[doublecol,superscriptaddress,floats,floatfix,showpacs]{epl2}
\usepackage{multirow,amssymb,amsbsy,amsmath}
\usepackage{graphicx}
\usepackage{verbatim}
\usepackage{booktabs}
\usepackage{dcolumn}
\usepackage{bbm}
\usepackage{dsfont}
\makeatletter
\DeclareMathOperator{\tr}{tr}

\usepackage{color}

\title{Non-Markovian quantum dynamics: What does it mean?}

\author{C.-F. Li \inst{1,2}$\footnote{email: cfli@ustc.edu.cn}$, G.-C. Guo \inst{1,2} \and
J. Piilo\inst{3}$\footnote{email: jyrki.piilo@utu.fi}$
}

\institute{
  \inst{1} CAS Key Laboratory of Quantum Information, University of Science and Technology of China, Hefei 230026, China\\
  \inst{2} CAS Center For Excellence in Quantum Information and Quantum Physics, University of Science and Technology of China, Hefei 230026, China\\
  \inst{3} Turku Centre for Quantum Physics, Department of Physics and Astronomy, University of Turku, FI-20014 Turun yliopisto, Finland \\
 
  }

\date{\today}

\abstract{
During the last ten years, the studies on non-Markovian open system dynamics has become increasingly popular and having contributions from diverse set of research communities. This interest has arisen due to fundamental problematics how to define and quantify memory effects in the quantum domain, how to exploit and develop applications based on them, and also due to the question what are the ultimate limits for controlling open system dynamics. We give here a simple theoretical introduction to the basic approaches to define and quantify quantum non-Markovianity -- also highlighting their connections and differences. 
In addition to the importance of the development for open quantum systems studies, we also discuss the implications of the progress for other fields including, e.g., formal studies of stochastic processes and quantum information science, and conclude with possible future directions the recent developments open.
}


\begin{document}
\maketitle

\section{Introduction}
%Everything what happens now depends, in some sense, from the past events (without Big Bang, I would not write this now). However, whether memory of the past events exist at current point of time, or if existing whether influences what happens now, is another matter. In other words, do memories of the past events influence the events now? This is rather complicated problem, and even more so when considering quantum dynamics.
 
The research on quantum dynamics -- which describes how quantum systems evolve over time --  is vital to our understanding of quantum physics and microscopic phenomena both in nature and in controlled experiments. 
In general, solving the dynamics gets increasingly difficult when the quantum system has a large number of degrees of freedom or when it is composed of several interacting subsystems. However, in this case we are often interested in how the dynamics of one of the subsystems, or the combination of some of them -- the open system --  is influenced by the interaction with the remaining subsystems or degrees of freedom -- the environment. Thereby, we do not necessarily need to solve  the complete unitary dynamics of the total system, composed of both the open system and its environment, but instead we can ask how the system-environment interaction influences the dynamics of the open system only. 

This leads to the field of open quantum systems~\cite{breuer2007,weiss2012,rivas2012} and also means that the dynamics of an open system is in general non-unitary leading to decoherence and loss of various quantum properties of the open system with time.  Thereby, understanding decoherence and open system dynamics is interesting for fundamental reasons and crucial for practical applications of quantum physics which rely in our ability to preserve quantum properties of open systems. In general and most strictly speaking, all realistic quantum systems interact with their environments and thereby must be considered as open systems.

The state of open quantum system is commonly described by density operator (or density matrix) and in general there is no unique equation of motion for its evolution. The specific mathematical form may depend, e.g.,  on the used techniques and approximations done in its derivation. Moreover, there exists also a large number of stochastic methods where the density operator of the open system is obtained as an average over the generated  ensemble of pure state evolutions~\cite{Plenio98,Percival,Garraway1997,Breuer99,Strunz1999,Piilo08}. However, for open system density operators there exists an equation of motion which represents perhaps the most important result in the theory of open systems to date. This is the Gorini-Kossakowski-Sudarshan-Lindblad (GKSL) master equation~\cite{Gorini,Lindblad,history}
\begin{eqnarray}
\label{eq:gksl}
\dfrac{d \rho_S(t)}{dt}&=&-i[H,\rho_S(t)]  \nonumber \\
&+ &\sum_k \gamma_k \left[ C_k\rho_S(t) C_k^\dagger-\dfrac{1}{2} \left \{ C_k^\dagger C_k,\rho_S(t) \right \} \right].
\end{eqnarray}
%%%%%%%%%%%%%%%%%%%%%%%%%%%%%%%%%%%%%%%%
Here, $\rho_S$ is the density operator of the open system while Hamiltonian $H$ generates its unitary dynamics.
$\gamma_k$ are positive constant rates, and $C_k$ are the jump operators with $k$ indexing the different decoherence channels.
We have also used $ \hbar =1$.

There is direct connection between the form of this equation and the semigroup property of the corresponding family of dynamical maps $\Phi_{t,0}$ parametrised by time $t$.
In general, the dynamical map is introduced via relation
$\rho_S(0) \rightarrow\rho_S(t) = \Phi_{t,0} \rho_S(0)= \tr_E [U(t) \rho_S(0) \otimes \rho_E U^{\dagger}(t)]$. Here, $E$ refers to the environment of the open system and $U(t)$ is the time evolution operator corresponding to the total system Hamiltonian $H_S+H_E+H_{SE}$ where $H_{SE}$ contains the system-environment interaction.
The semigroup property of the map, in turn, corresponds to the condition $\Phi_{t_1+t_2,0} = \Phi_{t_2,0}  \Phi_{t_1,0}$. 
Yet another important property of the dynamical map is complete positivity (CP), i.e., map $\Phi_{t,0}$ is CP when it fulfils the condition $(\Phi_{t,0}\otimes \mathcal{I}_{d_S}) \rho_{SA} \geqslant 0$. Here, $\mathcal{I}_{d_S}$ is the identity map arising from ancillary Hilbert space having the same dimension $d_S$ as the open system Hilbert space and 
$\rho_{SA}$ is the density operator for the combined state of the open system and ancilla. This guarantees that the eigenvalues of the density operator remain non-negative when including possible ancillary systems.
Thereby, an important feature of the GKSL equation \eqref{eq:gksl} is that it guarantees the CP property of the dynamical map and the corresponding open system time evolution.  In other words, when the CP dynamical map fulfils the semigroup property, the master equation is of the form of (1), and this guarantees that physical and positive initial states of the open system also remain such during their time evolution. Note that CP is stronger condition than that of positivity (P) which corresponds to $\rho_S(t)=\Phi_{t,0}\rho_S(0)  \geqslant 0$.

Despite the usefulness and celebrated status of the GKSL master equation \eqref{eq:gksl}, there exists also many models and open system evolutions which go beyond what this equation and semigroup dynamics is able to describe. In this case, the decoherence rates $\gamma_k$, and possibly also jump operators $C_k$, become time-dependent, i.e., $\gamma_k \rightarrow \gamma_k(t)$ and $C_k \rightarrow C_k(t)$. Traditionally the GKSL master equation has also been used as a border line separating memoryless Markovian and non-Markovian regime. In general, the increase in the ability to control the open system dynamics and properties of the environment during the last two decades has stimulated the need to develop tools and understanding of quantum dynamics beyond GKSL equation.

The difference between Markovian and non-Markovian character of classical stochastic processes has a clear and rigorous formulation (see, e.g., Ref.~\cite{breuer2007} and references therein).
 However, if, how, and to which extent these results can be exploited  to define non-Markovianity in quantum domain and for open system dynamics is a highly subtle question.  As a matter of fact, open quantum systems can display very rich dynamical features, e.g.,  re-coherence (recovery of earlier lost quantum properties), which do not have a direct classical counterpart. Therefore, it is useful not only to use the previous knowledge about non-Markovianity in classical domain but also to look for ways to exploit quantum features of open system dynamics when defining and quantifying quantum non-Markovianity. Considering memory effects in some sense indicating past states influencing the changes of the state at the current point of time, seems a plausible approach. However, when scrutinising this approach more rigorously, it is not obvious how to define Markovian -- non-Markovian border for quantum dynamics.
 
Indeed, during the last decade, a large number of different definitions and quantifiers of non-Markovian memory effects in the quantum domain have appeared in literature, see, e.g.,~\cite{wolf,NMprl,rivas,fisher,cor1,geo,cano,ff,channelcap,ipow1,illu,pine,zhe,CMM2016}. The debate about their features, connections, and usability has been intense~\cite{HPjphysb,RevRivas,RevRMP,LiRev} -- to say the least. Generally speaking, there are two early major  lines of research here: (i) focus on the mathematical properties of the dynamical map; (ii) a physically oriented approach independently whether the dynamical map is known or not. In former case, the starting point is the CP-divisibility property of the dynamical map~\cite{rivas} -- and in the latter case the focus is on the concept of information flow between the open system and its environment~\cite{NMprl}. In general, the two lines were initiated as separate directions of enquiry and do not coincide. However, during the last few years their relations and connecting interpretations have become increasingly known including also attempts to develop a  general hierarchy for non-Markovian features of open system dynamics and for methods used for this purpose~\cite{LiRev}. 
When going beyond information flow or divisibility properties, one can use also several other characterisations and concepts for describing memory effects in open system dynamics. These include, e.g., concepts of Fisher information~\cite{fisher}, correlations~\cite{cor1}, Bloch volume~\cite{geo}, negativity of the decay rates~\cite{cano}, channel capacities~\cite{channelcap}, interferometric power~\cite{ipow1}, Gaussian channels~\cite{illu}, and spectra of dynamical maps~\cite{CMM2016}.

This Perspective Article focusses on some theoretical developments of defining and quantifying non-Markovianity whilst applications and experiments are discussed in the accompanying Perspective Article~\cite{p2}.
These  articles can be used as short primers for recently published extensive reviews on non-Markovian quantum dynamics~\cite{HPjphysb,RevRivas,RevRMP,LiRev,devega} and for research literature dealing with this fascinating area of modern quantum physics.
 
\section{System-environment information flow}
It has been perhaps well-known for long time, that in general the open system and its environment exchange information in addition to 
energy~\cite{zurek}.
One of the early works that discussed  the system-environment information exchange in the context of non-Markovian open system dynamics is a stochastic non-Markovian quantum jump (NMQJ) method~\cite{Piilo08}. 
The open system density operator $\rho_S$ is obtained as an ensemble average over stochastic pure state evolutions, $\rho_S(t)= \sum_{\alpha}\frac{N_{\alpha}(t)}{N} 
|\psi_{\alpha} (t) \rangle \langle\psi_{\alpha} (t)|$. Here, $N_{\alpha}(t)$ is the number of ensemble members in the state $|\psi_{\alpha}(t) \rangle$ in the total ensemble of size $N$ and each pure state realization contains randomly occurring quantum jumps.
The key feature of the method is the appearance of reverse jumps in non-Markovian region which - instead of destroying quantum coherences - restore coherence (re-coherence) describing the reverse information flow from the environment back to the system.  
For example, for a undriven two-level atom interacting with electromagnetic environment in its ground state, the atom may emit a spontaneous photon in Markovian regime destroying superposition $c_g |g\rangle + c_e |e\rangle \rightarrow  |g\rangle$, where $c_g$ and $c_e$ are the probability amplitudes prior the quantum jump.
 However, when the electromagnetic environment has non-trivial structure and spectral density, in the corresponding non-Markovian regime, the two-level atom may recreate its earlier lost superposition via reverse jump  in a given realization  taking the atom from the ground state back to a superposition state $ |g\rangle \rightarrow c_g' |g\rangle + c_e' |e\rangle$. This shows up in the ensemble average over all realizations and in the open system density matrix $\rho_S$ as non-monotonic behaviour of the excited state population and magnitude of coherences.
This gives insight into the memory effects but does not quantify nor define non-Markovianity. 

In principle and \textit{a priori}, there are a number of choices for rigorous quantification of the information flow depending what one means by the word ``information''. It is also worth keeping in mind that for open system both the system-environment correlations and changes in the environmental state influence how open system evolves and how information flow should be understood~\cite{eplse,RevRMP}.
One of the first attempts to quantify system-environment information flow was based on the concept of trace distance~\cite{NMprl}
\begin{equation}
\label{eq:D}
 D(\rho^1,\rho^2) = \frac{1}{2} ||\rho^1-\rho^2||_1.
\end{equation}
Here, $\rho^1$ and  $\rho^2$ are two density operators and the trace norm for trace class operator $A$ is defined via $||A||_1 = {\mathrm{tr}}|A|$ with the modulus of $A$ given by $|A|=\sqrt{A^{\dagger}A}$.
Trace distance is invariant for unitary dynamics and contraction for CP-dynamical maps, i.e., given two initial open system states 
$\rho_S^1(0)$ and  $\rho_S^2(0)$, the trace distance between the time evolved states never exceeds its initial value
 $D[\rho_S^1(t),\rho_S^2(t)] \leq D[\rho_S^1(0),\rho_S^2(0)]$.

 In terms of the concept of information, trace distance $D$ gives the maximum probability  to distinguish two quantum states in a single shot experiment -- the probability being equal to $\frac{1}{2}[1+D(\rho^1,\rho^2)]$.
 Therefore, with increasing $D$, we also have more information which one of the two alternative states we have. 
 Even though $D$ can not increase under CP maps, and hence under the dynamical map $\Phi_{t,0}$,
 this does not mean that trace distance behaves in monotonic way as a function of time. 
 Therefore, we can quantify the degree of non-Markovianity and information backflow by calculating how much the trace distance - and thereby the distinguishability - increases during the whole time evolution of the open system. To obtain a single number, we maximise over all initial pairs of states and define the measure $\mathcal{N}_{BLP}$ as
\begin{equation}
\label{eq:N}
 {\mathcal{N}_{BLP}} = \max_{\rho_S^{1,2}(0)} \int_{\dot{D}(t)>0}
  dt \; \dot{D}(t)
\end{equation}
where $\dot{D}(t)=  \frac{d}{dt}D[\rho_S^1(t),\rho_S^2(t)]$.

For simple qubit open systems with limited number of decoherence channels, this measure is often straightforward to calculate. 
For example, a two-level atom undergoing amplitude damping with jump operator $\sigma_- = |g\rangle\langle e|$ and time-dependent decay rate $\gamma(t)$ in the master equation corresponding to Eq.\eqref{eq:gksl}, the optimizing initial pair is composed of the ground state $|g\rangle$ and excited state $|e\rangle$. Moreover, for this pair of states there is a direct relationship between the change of the trace distance and the sign of the decay rate given by
$\dot{D}(t)=-\gamma(t)\exp[-\Gamma(t)]$ where $\Gamma(t)=\int_0^t dt' \gamma(t')$. 
For further simple examples, see, e.g., the recent Colloquium article~\cite{RevRMP}.
 In general, the sign of the decay rate, in particular when the rates become negative, can be often associated to the backflow of information though in general the question is a subtle one~\cite{jphysb2012,cano,ninasrep}.  With increasing size of the open system's Hilbert space, the optimization task becomes increasingly difficult and one has to often use numerical techniques or random sampling of states,  and possibly also restrict the study to specific set of initial states. 
 For further simplifications how to find the maximizing pair, see~\cite{sa1,sa2}.
 %Note however, that there exists further refinements which show that the maximizing pair need to have orthogonal support~\cite{sa1} and the concept of direction in the state space can be exploited when performing the maximisation~\cite{sa2}. 
 In general, the trace distance measure has become quite common when dealing with non-Markovian dynamics and has been also used in various applications and experiments, see, e.g.,~\cite{NMNP,nmepl,telep,sdepl,nvnm,DJ}.

\section{Divisibility of the dynamical map}
The divisibility property of the dynamical map characterises essentially whether a given map can be divided into two or more legitimate maps fulfilling a given criteria. Consider the following concatenation of a CP dynamical map  $\Phi_{t_2,0}=\Phi_{t_2,t_1}\Phi_{t_1,0}$ with $t_2\geqslant t_1 \geqslant 0$.
If the dynamical map from point of time $t_1$ to $t_2$, $\Phi_{t_2,t_1}$, is not CP (not P), then the original map from time $0$ to $t_2$, $\Phi_{t_2,0}$, is not CP-divisible (not P-divisible). In other words, if the original legitimate map cannot be obtained as concatenation of two legitimate maps with given criteria, then the original map is called non-divisible.
 
In related contexts, the mathematical divisibility properties of quantum channels have been studied already in Ref.~\cite{WC2008}. 
Soon after this, a measure for non-Markovianity by using the concept of CP-divisibility for dynamical maps was developed~\cite{rivas}.  The starting point here is the following definition considering the points of time $t$ and $t+\epsilon$
\begin{equation}
\label{eq:nCP}
f(t+\epsilon,t)=||\left(\Phi_{(t+\epsilon,t)}\otimes\mathcal{I}_{d_S}\right)\left(|\Psi\rangle_{SA}~_{SA}\langle\Psi | \right)||_1.
\end{equation}
Here, $|\Psi\rangle_{SA}$ is a maximally entangled state between the open system $S$ and ancillary system $A$ whilst $f(t+\epsilon,t)=1$ for a CP map and  $f(t+\epsilon,t)>1$ for a non-CP map.
By using infinitesimal $\epsilon$, one can then define $g(t)$ as
\begin{equation}
g(t)=\lim_{\epsilon\rightarrow0^+}\frac{f(t+\epsilon,t)-1}{\epsilon}
\end{equation}
and integrate this over time
\begin{equation}
\label{eq: RHP}
\mathcal{I}_{RHP}=\int_0^\infty g(t)dt
\end{equation}
to quantify non-CP divisible character of the dynamical map and non-Markovianity in this sense.
%Note that this number in general is independent of the used initial state of the open system. Therefore it fully describes a mathematical property of the dynamical map $\Phi$ whilst in principle some initial open system states may follow Markovian evolution while some others non-Markovian one for a given map. 
As for the trace distance measure $\mathcal{N}_{BLP}$, the divisibility measure $\mathcal{I}_{RHP}$ can be calculated in quite straightforward manner for simple qubit systems. For example, a qubit undergoing dephasing with $\sigma_z$ as a jump operator and time dependent rate $\gamma(t)$  in the master equation corresponding to Eq.\eqref{eq:gksl}, the quantifier obtains the value $\mathcal{I}_{RHP}=-2\int_{\gamma(t)<0}\gamma(t)dt$.

\section{Relation between information flow and divisibility}

In general, the non-Markovianity measures based on the CP-divisibilty property and the trace distance  dynamics  do not coincide.
If  there is information backflow in terms of trace distance, then the map breaks CP-divisibility. However, the converse does not always happen.
The two measures coincide for single qubit and single decoherence channel open system dynamics but otherwise their relation is quite a subtle issue. 
For example, there exist maps, which break CP-divisibility for all points of time, while the dynamics is still P-divisible, trace distance decreasing in monotonic manner and with the possibility to simulate them with classical Markovian process~\cite{cano,ninasrep}.  
Further work has demonstrated interesting relations between information flow and divisibility measures. However, this requires the use of ancillary systems or some additional prior information, or both of them.

Trace distance $D$ is based on equal probabilities of preparing the two states, i.e., the preparation is uniformly random and there is no prior additional  information which one of the two states is prepared. However, additional information, which is not initially contained in the open system, can be helpful for further modifications. In particular, one can consider Helstrom matrix $\Delta$
\begin{equation}
\label{eq:helstrom}
\Delta=p_1\rho^1-p_2\rho^2,
\end{equation}
where $p_1$ and $p_2$ are the prior probabilities of the corresponding states. The information interpretation with one shot two state discrimination problem is also valid here~\cite{CKR2011}.
Consider now two time evolved  states 1 and 2 with ancillary system
\begin{equation}
  \rho_{SA}^{1,2}(t) = [ \Phi_{t,0}\otimes\mathcal{I}_{d_S} ]\rho_{SA}^{1,2}(0).
\end{equation}
Now, it is possible to show that the trace norm $E(t)$ of the Helstrom matrix $\Delta$, $E(t)=||\Delta(t)||_1=||p_1\rho_{SA}^{1}(t) -  p_2\rho_{SA}^{2}(t) ||_1$,
monotonically decreases  if and only if the map  $\Phi_{t,0}$ is CP-divisible~\cite{CKR2011}. This is valid for bijective maps, for generalization see~\cite{crs}. Therefore, one can also consider the case 
$\frac{d}{dt} E(t)>0$
in an analogous manner, compared to the trace distance measure, as a quantifier for non-Markovianity. 
This demonstrates that CP-divisibility measure has information flow interpretation when using ancillary systems and prior information about the preparation probabilities of the states.

One can take a step further and instead of considering only two states in the discrimination problem, ask how to minimise the discrimination error in the ensemble $\mathcal{E}=\{ p_i, \rho^i \}$ of quantum states with arbitrary and finite ensemble size $N$~\cite{BD2016}. Here, $p_i$ is the prior probability to have state $\rho^i$. This leads to the concept of guessing probability
\begin{equation}
\label{eq:guessing}
P_g(\mathcal{E})=\max \sum_i p_i\tr [P^i \rho^i ],
\end{equation}
where the maximisation is over all Positive Operator Valued Measures (POVMs) $P_i$
and leads to the maximum average probability for correct guesses about the states.
Consider now the combined Hilbert space of the open system space with an ancillary space given by 
$\mathcal{H}_S\otimes\mathcal{H}_A$.
The corresponding time evolution of the joint state $\rho_{SA}$ is given by
$\rho_{SA}(t)=(\Phi_{t,0}\otimes\mathcal{I}_{d_A})\rho_{SA}(0)$.
Now it is possible to show~\cite{BD2016}, that the dynamical map $\Phi_{t,0}$ breaks the CP-divisibility if and only if there exists auxiliary Hilbert space $\mathcal{H}_{A}$,  finite ensemble of time evolved bipartite states $\mathcal{E}_t=\{ p_i, (\Phi_{t,0}\otimes\mathcal{I}_{d_A})\rho_{SA}^{i}(0) \}$ and discrete times $t_2 >t_1$ so that
\begin{equation}
\label{eq:bd}
P_g(\mathcal{E}_{t_2}) >  P_g(\mathcal{E}_{t_1}). 
\end{equation}
In other words, if the above condition holds, then the information flow interpretation for CP-divisibility is provided this time via temporarily increasing guessing probability which also has a connection to data processing inequalities~\cite{BD2016,bus}. This result is  satisfying from mathematical point of view and general from information theoretical point of view. However, optimising over the type of auxiliary Hilbert space, probability distributions and POVMs  seem to make it difficult for practical purposes.

Let us now take a step back, and ask whether it is possible to find information flow interpretation for CP-divisibility when considering only two states, instead of the ensemble of states, and using trace distance $D$ with uniformly random choice of states, instead of a prior info used by Helstrom matrix $\Delta$. 
This is indeed possible for bijective maps and when using ancillary system which has a dimension $d_S+1$ when open system  has dimension $d_S$~\cite{BJA2017}. In another words, the dynamical map $\Phi_{t,0}$ is CP-divisible if and only if the trace distance $D$ decreases or remains constant as a function of time for all pairs of initial system-ancilla states. In mathematical form this can be expressed with times $t_2>t_1$
as
\begin{eqnarray}
D[(\Phi_{t_2,0}\otimes \mathcal{I}_{d_S+1})\rho_{SA}^1(0), (\Phi_{t_2,0}\otimes \mathcal{I}_{d_S+1})\rho_{SA}^2(0)]
 \nonumber \\
\leqslant
D[(\Phi_{t_1,0}\otimes \mathcal{I}_{d_S+1})\rho_{SA}^1(0), (\Phi_{t_1,0}\otimes \mathcal{I}_{d_S+1})\rho_{SA}^2)(0)].
\end{eqnarray}
Therefore, when this inequality is broken, the dynamical map is not CP-divisible, and there is information backflow when using extended system with $d_S+1$ dimensional ancilla~\cite{BJA2017} -- for results beyond bijective maps, see~\cite{crs,chakra}. 
%Interestingly, 
%it is also possible to show how to construct a pair of separable system ancilla states which display the information backflow in this case~\cite{BJA2017}. 
 
Let us consider now the case where no ancillas are used and this time using the Helstrom matrix $\Delta$ instead of the trace distance $D$. Having two initial open system states, $\rho_S^1$ and $\rho_S^2$, with corresponding prior probabilities $p_1$ and $p_2$,
 it is then possible to prove for bijective maps that the map is P-divisible if and and only if the trace norm $E$ of the Helstrom matrix with evolved states 
\begin{equation}
E(t)=|| \Delta(t)||_1=|| p_1\Phi_{t,0}(\rho_S^1) - p_2\Phi_{t,0}(\rho_S^2)||_1
\end{equation}
decreases monotonically~\cite{gd}, i.e., $\dot{E}(t) = \frac{dE(t)}{dt}\leqslant 0$ -- for results beyond bijective maps, see~\cite{crs}.
This means that when  $\dot{E} >0$, P-divisibility is broken, and  one has now information flow interpretation for P-divisibility provided that one has prior information about the probability of the two initial states. 
%Note also that trace distance is invariant in uniform shift of the two states, which can break the P-divisibility of the map. Therefore the use of Helstrom matrix helps in detecting this effect -- though it is a subtle issue how one could implement in the laboratory these non-unitary and non-positive operations.

Table 1 collects the basic features of all the quantifiers described above.
It is also worth noting that it is possible to generalize previously mentioned CP and P-divisibility properties to the concept of 
k-divisibility and use this to quantify the degree of non-Markovianity~\cite{mani}. This may become useful when considering open system Hilbert spaces which have dimension $d_S>2$.

\begin{table}[t]

\begin{tabular}{llllll}
   \hline
         Ref.
       & Quantifier
       & Prior 
       & Ancillas
       & Information \\
       & 
       & info
       & 
       &  flow \\

       \hline
      \cite{NMprl}   & D  & no   & no  & D \\
       \cite{rivas}  & CP-div  & no   & $d_S$ & -  \\
      \cite{CKR2011} & CP-div  & yes   & $d_S$  & $E$   \\     
      \cite{BD2016} & CP-div & yes   & $d_S$  & $P_g$   \\
     \cite{BJA2017}  & CP-div & no  & $d_S+1$  & D    \\
      \cite{gd}   & P-div & yes  & no & $E$  \\
     \end{tabular}
     \caption[t2]{\label{tab:div-info}
The basic features of connecting information flow and divisibility. Here, $D$ denotes trace distance, $E$ the trace norm of the Helstrom matrix $\Delta$, $P_g$ guessing probability, CP-div (P-div) indicates CP (P) divisibility. For those using ancillas, the dimensionality of the corresponding Hilbert space is mentioned when $d_S$ is the dimension of the open system Hilbert space. 
}
\end{table}

\section{Classical vs.~quantum stochastic processes}

In addition to the problematics how to define and quantify memory effects in open quantum system dynamics described by density operator evolutions, it is worthwhile to ask what is the relationship of non-Markovianity between classical and quantum stochastic processes. Take a classical stochastic process where the random variable can take a value from the set $\{x_i\}$ and consider different points of time $t_n\geqslant t_{n-1}\geqslant ... \geqslant t_1 \geqslant t_0$. The corresponding stochastic process is Markovian if the following equation holds for the associated conditional transition probabilites
\begin{equation}
\label{c-pr}
P(x_n,t_n | x_{n-1}, t_{n-1}; ... ; x_0, t_0) = P(x_n,t_n | x_{n-1}, t_{n-1}).
\end{equation}
In other words, the transition probability for the current value depends only on the most recent value of the random variable and is independent from all the other previous points of time and values. It is not obvious, if and how this property of the process can be transferred or generalized to quantum realm since for quantum processes  measurements influence the state of the quantum system and their evolution. 
%Moreover, it is possible to choose a measurement scheme for a given quantum system  in a multitude of different ways.

One of the early quantifiers of quantum non-Markovianity -- the loss of CP-divisibility property~\cite{rivas} -- can be considered analogous to classical definition in the following sense~\cite{RevRivas,basnjp}. Consider  a classical process with one-time probability $P(x,t)$ and linear map (transition matrix) $T$ connecting the values and probabilities at two different points of time as
$P(x_1,t_1)=\sum_{x_0} T(x_1,t_1|x_0,t_0) P(x_0,t_0)$. Stochastic process can be defined to be divisible when $T$ fullfils the following relations: i) $\sum_{x_2}  T(x_2,t_2|x_1,t_1)=1$; ii)   $T(x_1,t_1|x_0,t_0) \geqslant 0$; iii)  $T(x_3,t_3|x_1,t_1) =  \sum_{x_2}  T(x_3,t_3|x_2,t_2) T(x_2,t_2|x_1,t_1)$ for all points of time $t_3\geqslant t_{2}\geqslant  t_1 \geqslant t_0$. However, note that there exists divisible processes which are non-Markovian.
Going for quantum case, one replaces now the divisibility property of the transition matrix $T$ with the divisibility of the dynamical map $\Phi_{t,0}$. 
In particular, considering CP-divisibility, one checks when the map $\Phi_{t_2,t_1}$ becomes non-CP in the concatenation $\Phi_{t_2,0}= \Phi_{t_2,t_1}\Phi_{t_1,0}$ for $t_2\geqslant t_1 \geqslant 0 $.
Therefore, it is reasonable to think the loss of CP-divisibility as an analogous indicator of non-Markovianity when going from classical to quantum processes. 
However, it is worth keeping in mind the restriction of the analogy to one time probabilities only. Moreover, there is another subtle point involved for the quantum case related to the assumption that the density operator remains diagonal in the same basis over the evolution. For more details see~\cite{RevRivas}.
In similar spirit and restrictions, it is also possible to show that when the dynamical map is P-divisible, then one can write down a corresponding classical Markovian stochastic rate process~\cite{gd} -- also demonstrating a connection and analogy with the classical definition of non-Markovianity.

To develop a more general correspondence of non-Markovianity between classical and quantum processes, one needs to go beyond the traditional concept of a CP-dynamical map describing open system dynamics. This may also indicate the difference between the studies of expectation values and multi-time statistics. For the latter, it is also possible to quantify the violation of quantum regression theorem~\cite{gua}. 
For the correspondence to classical Markovianity, a recent series of papers~\cite{modi1,modi2,modi3} exploited the concepts of process tensor and causal break. Consider a sequence of times $t_0<t_1<,...,<t_{k-1}$ where at each point of time one applies a control operation (CP-map) $\mathcal{A}_j^{(r)}$ on the open system. Here, $j$ labels the point of time and $r$ one of a set of operations. The whole sequence of operations is denoted by $\mathbf{A}_{k-1:0}$. 
The  process is now characterised with process tensor $\mathcal{T}_{k:0}$ which maps the sequence of operations to the density operator at later time $\rho_k = \mathcal{T}_{k:0} [\mathbf{A}_{k-1:0}]$.
Suppose now a measurement is done on the open system at time $t_k$ recording its outcome $r$ and the corresponding positive operator being $\Pi^{(r)}_{k}$. 
After the measurement, the open system is prepared in a randomly chosen but known  state $P_{k}^{(s)}$ belonging to a set $\{P_k^{(s)}\}$. This is said to break the causal link for the open system between its past $t_j\leqslant t_k$ and future $t_l>t_k$ and describes the concept of causal break. The open system state in a later point of time can be formally described with a normalized state
$\rho_l= \rho_l(P_{k}^{(s)}| \Pi^{(r)}_{k}; \mathbf{A}_{k-1:0})$.
In  other words, this opens the possibility to check whether the state $\rho_l$ depends on its conditional argument, i.e., on the choice of control operations and choice of prior measurement. The claim of~\cite{modi1} is now that this state is consistent with conditional classical probability distributions -- and not limited to one time probabilities only. 

Thereby this allows to define a quantum stochastic process to be Markovian when the following holds:  $\rho_{l}(P_{k}^{(s)} |\Pi^{(r)}_{k};\mathbf{A}_{k-1:0}) = \rho_{l}(P_{k}^{(s)})$ for all control operations, measurements, preparations, and points of time.
This means that the current state of the open system depends only what the randomly chosen state was after the measurement and is independent of all the control operations prior the measurement. Subsequently,  one can then classify a quantum stochastic process being non-Markovian if and only if there exists two different controls which produce different open system states after the causal break at time $t_l$, i.e.
\begin{equation}
\rho_{l}(P_{k}^{(s)}|\Pi^{(r)}_{k};\mathbf{A}_{k-1;0}) \ne \rho_{l}(P_{k}^{(s)}|\Pi^{'(r')}_{k};\mathbf{A}'_{k-1;0}).
\end{equation}
This criterion was given in reference~\cite{modi1}. Note also earlier works~\cite{lin79,acc82}, and the use of the  process matrix formalism~\cite{proma1,proma2} for non-Markovian studies. In~\cite{modi1}, it is also stated that all time-independent system-environment Hamiltonians produce non-Markovian open system evolution according to the criteria above when considering more than two time steps. Thereby, most, if not all, commonly used theoretical microscopic system-environment models presented in the earlier literature should display memory effects in the open systems evolutions.
This include cases where the exact open system dynamics -- without being disturbed by control operations and measurements -- follows the GKSL master 
equation~\eqref{eq:gksl} and the corresponding dynamical map has the semigroup property~\cite{gua}.

At this point, it is legitimate to ask, whether e.g. semigroup dynamics -- {\textit{per se}} and in itself -- carries memory effects or not?
One possible answer here is that if one considers the control operations as probes, then the answer may be positive.
However, the probe is disturbing and modifying the quantum dynamics. Thereby another answer may be that semigroup dynamics itself does not carry memory effects but it is the combination of this with probe and measurement modified dynamics which displays memory effects. Here, one could also conclude that it is the non-Markovian character of the system-environment interaction Hamiltonian combined with access to multi-time statistics which is being discussed, and not that of the dynamical map. 
It is also useful to keep in mind here the large scale hierarchy presented in~\cite{LiRev}.
 
\section{Conclusions and perspectives}
Even though open quantum systems have been studied for several decades, during the last ten years there has been a large amount of increasing activity in this area. This has been motivated by urge in understanding various dynamical features when going beyond most simple open system dynamics --  and for increasing understanding for their mathematical description when using and connecting several earlier developed concepts from different fields including mathematical foundations of quantum mechanics and quantum information theory.  The early developments and commonly used approaches include the concepts of information flow~\cite{NMprl} and divisibility~\cite{rivas}. By now, we have a large variety of quantifiers for non-Markovian quantum dynamics~\cite{fisher,cor1,geo,cano,ff,channelcap,ipow1,illu,pine,zhe,CMM2016,RevRivas,RevRMP,LiRev} and to different facets of memory effects which are also related to the way one is allowed to probe the open system.  For applications and experiments, see the accompanying Perspective article~\cite{p2}

The emphasis of the research seems to be turning from developing more definitions and quantifiers for non-Markovianity to understanding memory effects as a resource and how to combine the control of complex quantum systems with exploitation of memory effects. Indeed, discussion on full resource theory of non-Markovianity has began~\cite{rt1,rt2}, though not yet completed in similar manner as has been achieved in a number of other fields of quantum physics or concepts therein. 

Even though, several quantifiers of non-Markovianity are very general by definition, it is not always obvious how to use them when the dimensionality and complexity of the structure of open system increases. Thereby, there is a need to develop and find connections to, e.g., directly measurable observables which could be used in practical open systems to indicate the presence of memory effects. This would be very important when considering, e.g., many-body open quantum systems. It is also possible to consider and exploit recent developments on non-Markovianity in the contexts not usually considered in open system community. This could include, e.g., studying the concept of information flow when running a quantum algorithm~\cite{bae}. Moreover, interesting future directions also include problems on non-classical features and characterization of non-Markovian temporal processes~\cite{nc1,nc2}, which may be helpful when developing general resource theory of non-Markovianity.

In general, recent progress has been a fascinating and fruitful interplay between various formal mathematical descriptions and more practically motivated approaches allowing rapid developments, and we expect this to continue with implications beyond the traditional problems dealt by open system community.

\acknowledgments
This work was supported by the National Key Research and Development Program of China (No. 2017YFA0304100), the National Natural Science Foundation of China (Nos. 11774335, 11821404), Key Research Program of Frontier Sciences, CAS (No. QYZDY-SSW-SLH003), the Fundamental Research Funds for the Central Universities (No. WK2470000026), and Anhui Initiative in Quantum Information Technologies (AHY020100). We thank H.-P. Breuer, F.~Buscemi, D.~Chru\'sci\'nski, M.~J.~W. Hall, S.~Huelga, S. Maniscalco, K. Modi, \'A.~Rivas, A. Smirne, and B. Vacchini for discussions.


\begin{thebibliography}{xx}


\bibitem{breuer2007} 
H. P. Breuer and F. Petruccione, {\textit{The Theory of Open
Quantum Systems}} (Oxford University Press, Oxford,  2007).

\bibitem{weiss2012} 
U. Weiss, {\textit{Quantum Dissipative Systems}} (World Scientific, Singapore,  2012).

\bibitem{rivas2012}
\'A. Rivas and S. F. Huelga, {\textit{Open quantum Systems. An Introduction}} (Springer, Heidelberg, 2012).

\bibitem{Plenio98} 
M. B. Plenio and P. L. Knight, {\textit{Rev. Mod. Phys.}}, {\bf 70} (1998) 101.

\bibitem{Percival}
I. Percival, {\textit{Quantum State Diffusion}}  (Cambridge University Press, Cambridge,  2002).

\bibitem{Garraway1997} 
B. M. Garraway,   {\textit{Phys. Rev. A}}, {\bf 55},
(1997) 2290.

\bibitem{Breuer99}
H.-P. Breuer, B. Kappler, and F. Petruccione,  {\textit{Phys. Rev. A}} {\bf 59} (1999) 1633.

\bibitem{Strunz1999} W. T. Strunz, L.  Di\`{o}si, and N. Gisin,  {\textit{Phys. Rev. Lett.}}, {\bf 82} (1999).

\bibitem{Piilo08}
J. Piilo \textit{et al.},  {\textit{Phys. Rev. Lett.}}, {\bf 100} (2008) 180402.

\bibitem{Gorini} 
V. Gorini, A. Kossakowski, and E. C. G. Sudarshan,
 {\textit{J. Math. Phys.}}, {\textbf{17}} (1976) 821. 

\bibitem{Lindblad}
G. Lindblad,
 {\textit{Commun. Math. Phys.}}, {\textbf{48}} (1976) 119.

\bibitem{history}
D. Chru\'sci\'nski and S. Pascazio,
{\textit{Open Sys. Inf. Dyn.}}, {\textbf{24}} (2017) 1740001.

\bibitem{wolf} 
M. M. Wolf \textit{et al.},
%Assessing non-Markovian quantum dynamics. 
{\textit{Phys. Rev. Lett.}}, {\bf 101} (2008) 150402.

\bibitem{NMprl} 
H.-P. Breuer, E.-M. Laine and J. Piilo,
%Measure for the degree of non-Markovian behavior of quantum processes in open systems.
{\textit{Phys. Rev. Lett.}}, {\bf 103} (2009) 210401.

\bibitem{rivas} 
\'A. Rivas, S. F. Huelga and M. B. Plenio,
%Entanglement and non-Markovianity of quantum evolutions. 
{\textit{Phys. Rev. Lett.}}, {\bf 105} (2010) 050403.

\bibitem{fisher}
X.-M. Lu, X. Wang and C. P. Sun,
{\textit{Phys. Rev. A}}, {\bf 82} (2010) 042103.

\bibitem{cor1}
S. Luo, S. Fu and H. Song,
{\textit{Phys. Rev. A}}, {\bf 86} (2012) 044101.

\bibitem{geo}
S. Lorenzo, F. Plastina and M Paternostro,
{\textit{Phys. Rev. A}}, {\bf 88} (2013) 020102(R).

\bibitem{cano}
M. J. W. Hall \textit{et al.},
{\textit{Phys. Rev. A}}, {\bf 89} (2014) 042120.

\bibitem{ff}
F. F. Fanchini \textit{et al.},
{\textit{Phys. Rev. Lett.}}, {\bf 112} (2014) 210402.

\bibitem{channelcap}
B. Bylicka, D. Chru\'sci\'nski and S. Maniscalco,
%Non-Markovianity and reservoir memory of quantum channels: a quantum information theory perspective.
{\textit{Sci. Rep.}}, \textbf{4} (2014) 5720.

\bibitem{ipow1}
H. Shekhar \textit{et al.},
{\textit{Phys. Rev. A}}, {\bf 91} (2015) 032115.

\bibitem{illu}
G. Torre, W. Roga, and F. Illuminati,
{\textit{Phys. Rev. Lett.}}, {\bf 115} (2015) 070401.

\bibitem{pine}
C. Pineda \textit{et al.},
{\textit{Phys. Rev. A}}, {\bf 93} (2016) 022117.

\bibitem{zhe}
Z. He \textit{et al.},
{\textit{Phys. Rev. A}}, {\bf 96} (2017) 022106.

\bibitem{CMM2016}
D. Chru\'sci\'nski, C. Macchiavello and S. Maniscalco,
{\textit{Phys. Rev. Lett.}}, {\bf 118} (2017) 080404.

\bibitem{HPjphysb}
H.-P. Breuer,
{\textit{J. Phys. B: At. Mol. Opt. Phys.}}, {\bf 45} (2012) 154001.

\bibitem{RevRivas}
\'A. Rivas, S. F. Huelga \and M. B. Plenio,
%Quantum non-Markovianity: characterization, quantification and detection.
{\textit{Rep. Prog. Phys.}}, {\bf 77} (2014) 094001.

\bibitem{RevRMP}
H.-P. Breuer \textit{et al.},
%Non-Markovian dynamics in open quantum systems,
{\textit{Rev. Mod. Phys.}}, {\bf 88} (2016) 021002.

\bibitem{LiRev}
L. Li, M. J. W. Hall and H. M.  Wiseman,
{\textit{Phys. Rep.}}, {\bf 759} (2018) 1.


\bibitem{p2}
C.-F. Li, G.-C. Guo and J. Piilo,
"Non-Markovian quantum dynamics: What is it good for?"
{\textit{EPL (Europhys. Lett.)}}, \textbf{XX} (2019) XXX.
 
\bibitem{devega}
I. de Vega and D. Alonso,
{\textit{Rev. Mod. Phys.}}, {\bf 89} (2017) 015001.

\bibitem{zurek}
W. H. Zurek, 
In: Meystre, P. and  Scully, M. (eds),   {\textit{Quantum Optics, Experimental Gravity, and Measurement Theory}}, NATO Adv. Science Ins. Series, {\bf{94}} (Springer, Boston, 1983). 

\bibitem{eplse}
E.-M. Laine, J. Piilo, and H.-P. Breuer,
{\textit{EPL }}, {\bf 92} (2010) 60010.

\bibitem{jphysb2012}
E.-M. Laine, K. Luoma  J. Piilo,
{\textit{J. Phys. B: At. Mol. Opt. Phys.}}, {\bf 45} (2012) 154004.


\bibitem{ninasrep}
N. Megier \textit{et al.},
{\textit{Sci. Rep.}}, {\bf 7} (2017) 6379.

\bibitem{sa1}
S. Wissmann \textit{et al.},
{\textit{Phys. Rev. A}}, {\bf 86} (2012) 062108.

\bibitem{sa2}
B.-H. Liu \textit{et al.},
{\textit{Sci. Rep.}}, {\bf 4} (2014) 6327.

\bibitem{NMNP}
B.-H. Liu \textit{et al.},
%Experimental control of the transition from Markovian to non-Markovian dynamics of
%open quantum systems.
{\textit{Nature Phys.}}, {\bf 7} (2011) 931.

\bibitem{nmepl} 
J.-S. Tang \textit{et al.},
{\textit{EPL (Europhys. Lett.)}}, \textbf{97} (2012) 10002.

\bibitem{telep} 
E.-M. Laine, H.-P. Breuer \and J. Piilo,
%Nonlocal memory effects allow perfect teleportation with mixed states.
{\textit{Sci. Rep.}}, \textbf{4} (2014) 4620.

\bibitem{sdepl} 
B.-H. Liu \textit{et al.},
{\textit{EPL (Europhys. Lett.)}}, \textbf{114} (2016) 10005.

\bibitem{nvnm}
J.F. Haase \textit{et al.},
{\textit{Phys. Rev. Lett.}}, {\bf 121} (2018) 060401.

\bibitem{DJ}
Y. Dong, \textit{et al.},
{\textit{ npj Quantum Information}}, \textbf{4} (2018) 3.

\bibitem{WC2008}
M. M. Wolf and J. Ignacio Cirac,
{\textit{Commun. Math. Phys.}}, {\bf 279} (2008) 147.

\bibitem{CKR2011}
D. Chru\'sci\'nski, A. Kossakowski  and \'A. Rivas,
{\textit{Phys. Rev. A}}, {\bf 83} (2011) 052128.

\bibitem{crs}
 D. Chru\'sci\'nski, \'A. Rivas, and E. St{\o}rmer, 
{\textit{ Phys. Rev. Lett.}}, {\bf 121} (2018) 080407.


\bibitem{BD2016}
F. Buscemi \textit{et al.},
{\textit{Phys. Rev. A}}, {\bf 93} (2016) 012101.

\bibitem{bus}
F. Buscemi,
in: Ozawa M., et al. (eds),   {\textit{Reality and Measurement in Algebraic Quantum Theory.}} Springer Proc. in Math. and Stat., {\bf 261} 135 (Springer, Singapore, 2015). 

\bibitem{BJA2017}
B. Bylicka  \textit{et al.},
{\textit{Phys. Rev. Lett.}}, {\bf 118} (2017) 120501.

\bibitem{chakra}
S. Chakraborty \textit{et al.},
{\textit{Phys. Rev. A}}, {\bf 99}  (2019) 042105.


\bibitem{gd}
S. Wissmann, H.-P. Breuer and B. Vacchini,
{\textit{Phys. Rev. A}}, {\bf 92} (2015) 042108.

\bibitem{mani}
D. Chru\'sci\'nski and S. Maniscalco,
%Degree of non-Markovianity of quantum evolution.
{\textit{Phys. Rev. Lett.}}, {\bf 112} (2014) 120404.

\bibitem{basnjp}
B. Vacchini  \textit{et al.},
{\textit{New J. Phys.}},  {\bf 13} (2011) 093004.


\bibitem{modi1}
F. A. Pollock \textit{et al.},
{\textit{Phys. Rev. Lett.}}, {\bf 120} (2018) 040405.

\bibitem{modi2}
F. A. Pollock \textit{et al.},
{\textit{Phys. Rev. A}}, {\bf 97} (2018) 012127.

\bibitem{modi3}
S. Milz \textit{et al.},
{\textit{Phys. Rev. Lett.}}, {\bf 123} (2019) 040401.

\bibitem{gua}
G. Guarnieri, \textit{et al.},
{\textit{Phys. Rev. A}}, {\bf 90} (2014) 022110.

\bibitem{lin79}
G. Lindblad,
{\textit{Commun. Math. Phys.}}, {\bf 65} (1979) 281.

\bibitem{acc82}
L. Accardi \textit{et al.},
{\textit{Publ. RIMS Kyoto Univ.}}, {\bf 18} (1982) 97.


\bibitem{proma1}
F. Costa  \textit{et al.},
{\textit{New. J. Phys.}}, {\bf 18} (2016) 063032.


\bibitem{proma2}
C. Giarmatzi and  F. Costa,
arXiv:1811.03722.


\bibitem{rt1}
S. Bhattacharya \textit{et al.},
arXiv:1803.06881.

\bibitem{rt2}
G. D. Berk \textit{et al.},
arXiv:1907.07003.

\bibitem{bae}
S. S. Roy and J. Bae,
{\textit{Phys. Rev. A}},  {\bf 100} (2019) 032303.

\bibitem{nc1}
A. Smirne  \textit{et al.},
{\textit{Quantum Sci.~Technol.}}, {\bf 4} (2018) 01LT01.

\bibitem{nc2}
S. Milz \textit{et al.},
arXiv:1907.05807.
\end{thebibliography}
\end{document}